\documentclass[12pt,epsbox]{article}
\usepackage{latexsym}
\usepackage{psfig,amssymb,epsf}

\makeatletter
\@addtoreset{equation}{section}
\renewcommand{\theequation}{\thesection.\arabic{equation}}

\newcommand{\kb}{\bar{k}}

\newcommand{\Xb}{\bar{X}}

\newcommand{\bR}{{\bf R}}

\newcommand{\bS}{{\bf S}}

\newcommand{\cA}{{\cal A}}
\newcommand{\cJ}{{\cal J}}

\newcommand{\Kh}{\widehat{K}}
\newcommand{\Xt}{{\widetilde X}}
\newcommand{\ct}{{\widetilde c}}
\newcommand{\cJt}{{\widetilde {\cal J}}}
\newcommand{\vphi}{{\varphi}}

\newcommand{\Xtb}{\bar{\widetilde{X^2}}}

\newcommand{\nn}{\nonumber \\}

\newcommand{\be}{\begin{equation}} \newcommand{\ee}{\end{equation}}
\newcommand{\bea}{\begin{eqnarray}} \newcommand{\eea}{\end{eqnarray}}

\font\zfont = cmss10 
\newcommand{\ZZ}{\hbox{\zfont Z\kern-.4emZ}}

\ifx\TwoupWrites\UnDeFiNeD\else\target{\magstepminus1}{11.3in}{8.27in}
\source{\magstep0}{7.5in}{11.69in}\fi
\newfont{\fourteencp}{cmcsc10 scaled\magstep2}
\newfont{\titlefont}{cmbx10 scaled\magstep3}
\newfont{\authorfont}{cmcsc10 scaled\magstep1}
\newfont{\fourteenmib}{cmmib10 scaled\magstep2}
\skewchar\fourteenmib='177
\newfont{\elevenmib}{cmmib10 scaled\magstephalf}
\skewchar\elevenmib='177
\makeatletter
\newcommand\nonsequentialeqnum{
\@addtoreset{equation}{section}
\def\theequation{\arabic{section}.\arabic{equation}}}
\newif\ifp@bblock \p@bblocktrue
\newcommand\nopubblock{\p@bblockfalse}
\newcommand\topspace{\hrule height 0pt depth 0pt \vskip}
\newcommand\p@bblock{\begingroup \tabskip=\hsize minus \hsize
\baselineskip=1.5\ht\strutbox \topspace-2\baselineskip
\halign to\hsize{\strut ##\hfil\tabskip=0pt\crcr
\the\Pubnum\crcr\the\date\crcr}\endgroup}

\renewcommand\titlepage{\ifx\TwoupWrites\UnDeFiNeD\null
\vspace{-1.7cm}\fi
\vskip0.6cm
\ifp@bblock\p@bblock \else\hrule height 0pt \relax \fi}
\makeatother
\newtoks\date
\newtoks\Pubnum
\newtoks\pubnum
\newcommand{\frontpageskip}{\vspace{12pt plus .5fil minus 2pt}}
\renewcommand{\title}[1]{\frontpageskip
\begin{center}{\titlefont #1}\end{center}\par}
\renewcommand{\author}[1]{\frontpageskip\par\begin{center}
{\authorfont #1}\end{center}
\nobreak
}

\renewcommand{\thanks}[1]{\footnote{#1}}
\renewcommand{\abstract}{\par\frontpageskip\centerline{
\fourteencp Abstract}
\vspace{8pt plus 3pt minus 3pt}}

\begin{document}

\begin{titlepage}
\hfill
\vbox{
    \halign{#\hfil         \cr
           TAUP-2724-03 \cr
           hep-th/0304103  \cr
           } 
      }  
\vspace*{20mm}
\begin{center}
{\Large {\bf  The Effective Action and Geometry\\
of Closed $N=2$ Strings}\\} 
\vspace*{15mm}
{\sc Dan Gl\"uck}
\footnote{e-mail: {\tt gluckdan@post.tau.ac.il}},
{\sc Yaron Oz} 
\footnote{e-mail: {\tt yaronoz@post.tau.ac.il, Yaron.Oz@cern.ch}}
and {\sc Tadakatsu Sakai}
\footnote{e-mail: {\tt tsakai@post.tau.ac.il}}

\vspace*{1cm} 
{\it {Raymond and Beverly Sackler Faculty of Exact Sciences\\
School of Physics and Astronomy\\
Tel-Aviv University , Ramat-Aviv 69978, Israel}}\\

\end{center}

\begin{abstract}

$N=2$ closed strings have been recently 
divided in hep-th/0211147 to two T-dual families
denoted by $\alpha$ and
$\beta$.
In $(2,2)$ signature both families have one scalar in the spectrum.
The scalar in the $\beta$-string is known to be a deformation 
of the target space K\"ahler potential and the dynamics is that
of self-dual gravity.
In this paper we compute the effective action of the scalar
in the $\alpha$-string. The scalar is a deformation of a potential
that determines the metric, torsion and dilaton.
The scalar is free and the 
dynamics is that of a self-dual curvature with torsion. 
The result is in agreement with a $\sigma$-model computation of Hull.

\end{abstract}
\vskip 1cm

April 2003

\end{titlepage}

\setcounter{footnote}{0}

\newpage

\section{Introduction}

Closed $N=2$ strings \cite{ademollo;76} possess local ${\cal N}=2$
supersymmetry on the string worldsheet.
Critical $N=2$ strings have a four-dimensional target space.
The supersymmetric structure implies that the target space
has a complex structure. Therefore it must be of signature $(4,0)$
or $(2,2)$.
In $(4,0)$ signature there are no propagating degrees of freedom
in the $N=2$ string spectrum.
In $(2,2)$ signature there is only one massless scalar  in the
spectrum and
the infinite tower of massive excitations of the string is absent.
The effective action of this scalar has been computed in \cite{ov}
which suggested its interpretation as a deformation of the
target space  K\"ahler potential. 
It was  argued in \cite{ov} that the $N=2$ strings may
describe a quantum theory of self-dual Einstein gravity in four
dimensions.

In \cite{oz;n2} the
$N=2$
closed strings have been divided into two families denoted by
$\alpha$ and $\beta$.
Consider $N=2$ strings in a flat background.
In order to construct the $N=2$ string we need to gauge the $N=2$ 
superconformal algebra (SCA) on the worldsheet. 
More precisely we have two copies of the $N=2$
algebra to consider: the left and right sectors. 
The free field
representation of the (left) $N=2$ SCA
takes the form
\begin{eqnarray}
&&T=
-{1\over 2}\eta_{IJ}\partial x^I\partial x^J-{1\over 4}\,\eta_{IJ}
\left( \partial\psi^I\psi^J+\partial\psi^J\psi^I\right),\nn
&&J={i\over 2}\cJ_{IJ}^L\psi^I\psi^J,\nn
&&G^{\pm}={i\over 2}\left(
\eta_{IJ}\pm i\cJ_{IJ}^L\right)\psi^I\partial x^J.
\label{sca}
\end{eqnarray}
Here $I,J=1,...,4$ denote the indices of the target space in a real
basis.
The metric is given by
$\eta_{IJ}={\rm diag}(-1,-1,+1,+1)$.
${\cal J}^L_{IJ}$ is a K\"ahler form related to the complex structure
$\cJ^K_{~J}$
by $\cJ_{IJ}=\eta_{IK}\cJ^K_{~J}$, and 
the index $L$ refers to the left sector.\footnote{For more details see
next section and \cite{oz;n2}.}
Similarly, 
we have the SCA generators in the right sector with a
complex structure ${\cal J}^R$.
The (conventional) $N=2$ string denoted by $\beta$-string
in \cite{oz;n2} is defined by having
the same complex structure in the left and right sectors
${\cal J}^L={\cal J}^R$.
On the other hand, $N=2$ string denoted by
$\alpha$-string in  \cite{oz;n2}
has different complex structures in the left and
right sectors \footnote{For an earlier discussion see \cite{Gates:1988tn}.}.
In fact the $\beta$- and $\alpha$-strings define two families
of $N=2$ strings related by T-duality
\cite{oz;n2}.

In $(2,2)$ signature both families have one scalar in the spectrum.
The scalar in the $\beta$-string is, as noted above, a deformation 
of the target space K\"ahler potential and the dynamics is that
of self-dual gravity.
The aim of this paper is to  compute the exact effective action of the scalar
in the $\alpha$-string. The scalar is a deformation of a potential
that determines the target space metric, torsion and dilaton
\cite{GHR}.
The dynamics is that of a self-dual curvature with torsion
 \cite{Hull}. 

The paper is organized as follows. In section 2, we consider the 
worldsheet description  of $N=2$ strings. 
We present the $\sigma$-model Lagrangian description of 
the $\alpha$-string using
a chiral and a twisted chiral superfields and construct the vertex operators.
In section 3, we compute
the genus zero
three-point and four-point scattering amplitudes of the $N=2$ strings scalar.
In section 4 we show, based on sections 2 and 3,  
that the 
$\alpha$-string scalar is free and that the 
dynamics is that of a self-dual curvature with torsion.
This has been anticipated in \cite{Hull} based on 
$\sigma$-model and conformal anomaly analysis.
In section 5 we construct an example of a gravitational $\alpha$-string background, based
on the space transverse to NS5-branes.

\section{Worldsheet description  of $N=2$ strings}

In this section we will discuss in some detail the worldsheet
description of the $\beta$- and $\alpha$-strings.
We consider the $\sigma$-model Lagrangian and the vertex operators.

\subsection{Complex Structure and $N=2$ SCA}

In the following we will review some aspects
of the complex structures on $\bR^{2,2}$ that are relevant to
the generators of $N=2$ SCA.
In the real basis $x^I=(x^1,x^2,x^3,x^4)$, the metric is given by
$\eta_{IJ}={\rm diag}(-1,-1,+1,+1)$.
We define a complex structure
\begin{equation}
{\cal J}^I_{~J}=
\left(
\begin{array}{cc}
i\sigma_2 & 0 \\
0         & i\sigma_2 
\end{array}
\right) \ .
\end{equation}
In the complex basis
\begin{equation}
z^1={x^1+ix^2\over\sqrt{2}},~~z^2={x^3+ix^4\over\sqrt{2}},
\end{equation}
the metric reads $\eta_{i\bar{j}}={\rm diag}(-1,+1)$, $i,\bar{j} =1,2$.
In this basis, the complex structure ${\cal J}^I_{~J}$ is diagonal:
\begin{eqnarray}
&&\cJ(z^1)=-i z^1, ~~\cJ(\bar{z}^{\bar 1})=+i \bar{z}^{\bar 1} \ , \nn
&&\cJ(z^2)=-i z^2, ~~\cJ(\bar{z}^{\bar 2})=+i \bar{z}^{\bar 2} \ .
\end{eqnarray}
The K\"ahler form $\cJ_{IJ}=\eta_{IK}\cJ^K_{~J}$ is given in the real
basis
by
\begin{equation}
\cJ_{IJ}=
\left(
\begin{array}{cc}
-i\sigma_2 & 0 \\
0         & i\sigma_2 
\end{array}
\right) \ .
\end{equation}
For later reference, we define the quadratic form in momenta
\begin{equation}
\label{cij}
k_i^I\cJ_{IJ}k_j^J=ic_{ij} \ ,
\end{equation}
where in the complex basis
\begin{equation}
\label{cijn}
c_{ij} = k_i\cdot\bar{k}_j-\bar{k}_i\cdot k_j \ ,
\end{equation}
and $ k_i\cdot\bar{k}_j \equiv \eta_{m\bar{n}} k_i^m\bar{k}_j^{\bar{n}}$.
For the on-shell momenta $k_i,~i=1,2,3,4$ with 
$k_i^2=0,~k_1+k_2+k_3+k_4=0$,
$c_{ij}$ obey the identities \cite{ov}
\begin{eqnarray}
{c_{12}c_{34}\over s} + {c_{23}c_{41}\over t} = u,~~
{c_{21}c_{34}\over s} + {c_{13}c_{42}\over u} = t,~~
{c_{13}c_{24}\over u} + {c_{32}c_{41}\over t} = s \ ,
\label{id;c}
\end{eqnarray}
where $s=-k_1\cdot k_2 \equiv -(k_1\cdot\bar{k}_2+\bar{k}_1\cdot k_2),
~t=-k_2\cdot k_3,~u=-k_1\cdot k_3$.

We will also need a second complex (and  K\"ahler) structure,
which in the real basis take the form
\begin{equation}
\widetilde{{\cal J}}^I_{~J}=
\left(
\begin{array}{cc}
i\sigma_2 & 0 \\
0         & -i\sigma_2
\end{array}
\right),\quad
\widetilde{\cJ}_{IJ}=\eta_{IK}\widetilde{\cJ}^K_{~J}=
\left(
\begin{array}{cc}
-i\sigma_2 & 0 \\
0         & -i\sigma_2
\end{array}
\right) \ .
\end{equation}
In the complex basis, the complex structure is given by
\begin{eqnarray}
&&\cJt(z^1)=-i z^1, ~~\cJt(\bar{z}^{\bar 1})=+i \bar{z}^{\bar 1} \ , \nn
&&\cJt(z^2)=+i z^2, ~~\cJt(\bar{z}^{\bar 2})=-i \bar{z}^{\bar 2} \ .
\end{eqnarray}
The K\"ahler form reads
\begin{equation}
\cJt_{i\bar{j}}=-i{\bf 1},~~\cJt_{\bar{i}j}=+i{\bf 1} \ .
\end{equation}
We define also
\begin{equation}
\label{ctij}
\ct_{ij}=-ik^I_i\cJt_{IJ}k_j^J=-k_i^1\bar{k}_j^1-k_i^2\bar{k}_j^2
+\bar{k}_i^1k_j^1+\bar{k}_i^2k_j^2 \ ,
\end{equation}
which obeys an equation similar to (\ref{id;c}), for four on-shell momenta. 

The $\beta$-string in $\bR^{2,2}$ is defined by the choice
\footnote{More precisely, there is a parameter space of
  $\beta$-strings corresponding to the choice of the complex
  structures in the $N=2$ SCA, see \cite{oz;n2}. 
  The results that will be presented later remain valid for any
  choice of the complex structure $\cJ_L=\cJ_R$.}
\begin{equation}
\cJ^L_{IJ}=\cJ^R_{IJ}=\cJ_{IJ} \ ,
\end{equation}
in  (\ref{sca}), while 
for the $\alpha$-string 
\begin{equation}
\cJ^L_{IJ}=\cJ_{IJ},\quad \cJ^R_{IJ}=\widetilde{\cJ}_{IJ} \ .
\end{equation}
The two $N=2$ strings are related 
by a T-duality along a spatial direction.

\subsection{A $\sigma$-Model Lagrangian Description}

Here we will consider the two-dimensional $\sigma$-model 
Lagrangian description of the $N=2$ strings.
We begin by reviewing 
the ${\cal N}=(2,2)$ superfield formulation in Euclidean
two-dimensional space-time with the complex coordinates $z$ and $\bar{z}$ (see e.g. \cite{GHR}).
The complex 
 supercovariant derivatives for the left and right movers $D_{\pm}$
and $\bar{D}_{\pm}$
are defined by
\begin{equation}
D_{\pm}={\partial\over\partial\theta^{\pm}}+\theta^{\mp}\partial_z,
\quad
\bar{D}_{\pm}=-{\partial\over\partial\bar{\theta}^{\pm}}
              +\bar{\theta}^{\mp}\partial_{\bar{z}} \ ,
\end{equation}
where $\theta^{\pm}$ are the complex fermionic coordinates in the
superspace.
Denote $z^{\pm},\bar{z}^{\pm}$ by
\begin{equation}
z^{\pm}=z\pm\theta^2,\quad
\bar{z}^{\pm}=\bar{z}\pm\bar{\theta}^2 \ ,
\end{equation}
with $\theta^2=\theta^+\theta^-,~
\bar{\theta}^2=\bar{\theta}^-\bar{\theta}^+$.
It is easy to see that
\begin{equation}
D_{\pm}z^{\mp}=0 \ .
\end{equation}

A chiral superfield $X$ is defined by
\begin{equation}
D_+X=\bar{D}_+X=0 \ ,
\end{equation}
and an  anti-chiral superfield $\bar{X}$ by
\begin{equation}
D_-\bar{X}=\bar{D}_-\bar{X}=0 \ .
\end{equation}
In components we have 
\begin{eqnarray}
X&\!=\!&X(z^-,\bar{z}^-,\theta^-,\bar{\theta}^-) \nn
&\!=\!&x(z^-,\bar{z}^-)+\sqrt{2}\,\theta^-\psi_L(z^-,\bar{z}^-)
+\sqrt{2}\,\bar{\theta}^-\psi_R(z^-,\bar{z}^-)
+2\theta^-\bar{\theta}^-F(z^-,\bar{z}^-) \ ,
\end{eqnarray}
and
\begin{eqnarray}
\bar{X}&\!=\!&\bar{X}(z^+,\bar{z}^+,\theta^+,\bar{\theta}^+) \nn
&\!=\!&\bar{x}(z^+,\bar{z}^+)
+\sqrt{2}\,\theta^+\bar{\psi}_L(z^+,\bar{z}^+)
-\sqrt{2}\,\bar{\theta}^+\bar{\psi}_R(z^+,\bar{z}^+)
+2\theta^+\bar{\theta}^+F(z^+,\bar{z}^+) \ .
\end{eqnarray}

Since $\{D_{\pm},\bar{D}_{\pm}\}=0$, one can define in two dimensions 
twisted chiral superfields $\Xt$ by
\begin{equation}
D_+\Xt=\bar{D}_-\Xt=0 \ ,
\end{equation}
and 
an anti-twisted chiral $\bar{\Xt}$ superfield by
\begin{equation}
D_-\bar{\Xt}=\bar{D}_+\bar{\Xt}=0 \ .
\end{equation}
In components we have 
\begin{eqnarray}
\Xt&\!=\!&\Xt(z^-,\bar{z}^+,\theta^-,\bar{\theta}^+) \nn
&\!=\!&x(z^-,\bar{z}^+)+\sqrt{2}\,\theta^-\psi_L(z^-,\bar{z}^+)
+\sqrt{2}\,\bar{\theta}^+\psi_R(z^-,\bar{z}^+)
+2\theta^-\bar{\theta}^+F(z^-,\bar{z}^+) \ ,
\end{eqnarray}
and
\begin{eqnarray}
\bar{\Xt}&\!=\!&
\bar{\Xt}(z^+,\bar{z}^-,\theta^+,\bar{\theta}^-) \nn
&\!=\!&\bar{x}(z^+,\bar{z}^-)
+\sqrt{2}\,\theta^+\bar{\psi}_L(z^+,\bar{z}^-)
-\sqrt{2}\,\bar{\theta}^-\bar{\psi}_R(z^+,\bar{z}^-)
+2\theta^+\bar{\theta}^-F(z^+,\bar{z}^-) \ .
\end{eqnarray}

T-duality along $X+\Xb$ exchanges chiral (anti-chiral)
superfields and twisted (anti-twisted) chiral superfields.
In the complex structure language, the left movers complex structure
does not change but the right movers complex structure does
(${\cal J} \leftrightarrow \widetilde{\cJ}$).

The $N=2$ $\sigma$-model action
for the $\beta$-string in a flat background is 
\begin{equation}
S^0_{\beta}=\int d^4xd^4\theta K_0(X^1,\bar{X}^1,X^2,\bar{X}^2) \ ,
\end{equation}
where $X^i,i=1,2$ are chiral superfields and  
$K_0$ is the K\"ahler potential for flat $(2,2)$ space
\begin{equation}
K_0(X^1,\bar{X}^1,X^2,\bar{X}^2)=-X^1\bar{X}^1+X^2\bar{X}^2
~\sim -X^1\bar{X}^1+{1\over 2}(X^2+\bar{X}^2)^2 \ .
\label{k0;beta}
\end{equation}
In the last equivalence we used the freedom of adding holomorphic
and antiholomorphic terms to the  K\"ahler potential
without changing the metric $g_{i\bar{j}} = \partial_{X^i}
\partial_{\bar{X}^j}K_0$.

The $N=2$ $\sigma$-model action
for the $\alpha$-string in a flat background is 
\begin{equation}
\label{khat}
S^0_{\alpha}=\int d^4xd^4\theta 
\widehat{K}_0(X^1,\bar{X}^{1},\Xt^2,\Xtb) \ ,
\end{equation}
where $X^1$ is a chiral superfield and
$\Xt^2$ is a twisted chiral superfield.  
$\widehat{K}_0$ is the  Legendre transform of $K_0$
with respect to  $X^2 + \bar{X}^2$
\cite{oz;n2} (see \cite{Buscher:sk,Buscher:uw,Ivanov:ec}).
Thus, 
\begin{equation}
\Xt^2+\Xtb={\partial K_0\over\partial (X^2+\bar{X}^2)}
=X^2+\bar{X}^2 \ ,
\end{equation}
and 
\begin{equation}
\widehat{K}_0(X^1,\bar{X}^{1},\Xt^2,\Xtb)
=K_0-(X^2+\bar{X}^{2})(\Xt^2+\Xtb)
=-X^1\bar{X}^1-{1\over 2}(\Xt^2+\Xtb)^2 \ .
\label{k0;alpha}
\end{equation}
The target space metric, torsion and dilaton are encoded in the
function
$\widehat{K}_0$
\cite{GHR}
\begin{equation}
\label{lt}
\hat{g}_{1\bar{1}}=
{\partial^2\hat{K}_0\over \partial X^1\partial \bar{X}^1}=-1,
~~
\hat{g}_{2\bar{2}}=-{\partial^2\hat{K}_0\over \partial \Xt^2\partial \Xtb}
=+1 \ .
\end{equation}
In flat space the torsion and dilaton vanish and we have
\begin{equation}
B_{1\bar{2}} = {\partial^2\hat{K}_0\over \partial X^1\partial
  \Xtb} = 0,~~
B_{2\bar{1}} = {\partial^2\hat{K}_0\over \partial \Xt^2\partial
  \bar{X}^1} = 0 \ ,
\end{equation}
and
\begin{equation}
\label{dila}
\varphi = \frac{1}{2}\log \hat{g}_{2\bar{2}} = 0 \ .
\end{equation}
Note that in order to get the dilaton equation (\ref{dila}), the
dilaton term \cite{FT} should be added to the $\sigma$-model
action $S^0_{\alpha}$.

\subsection{Vertex Operators}

The vertex operators for the massless scalar in the
$\beta$- and $\alpha$-strings are given by
\begin{equation}
V_{\beta}(k)=e^{i(\kb^1 X^1+\kb^2 X^2+k^1 \Xb^1+k^2 \Xb^2)} \ ,
\end{equation}
and 
\begin{equation}
V_{\alpha}(k)=e^{i(\kb^1 X^1+\kb^2 \Xt^2+k^1 \Xb^1+k^2 \Xtb)} 
\ ,
\label{vertex;alpha}
\end{equation}
respectively. The ghost part will be treated separately.
The vertex operator in the $(-1,-1)$-picture 
comes from the lowest component of the superfields
and is the same for the $\beta$- and $\alpha$-strings
\begin{equation}
V_L^{(-1,-1)}(z)=e^{ik\cdot x_L(z)},\quad
V_R^{(-1,-1)}(\bar{z})=e^{ik\cdot x_R(\bar{z})} \ ,
\end{equation}
with $k^2=0$.
Note that we use $x$ as the lowest component of the superfield $X$.

In order to obtain the vertex operators of higher superconformal ghost
numbers we need to use the picture-changing operations \cite{fms} (see also \cite{Li:1992rr}), 
that are
implemented by acting on the vertex operators with
the worldsheet supercharges given in (\ref{sca}).
Using the OPE's of the free fields of the form
\begin{equation}
x_L^I(z)x_L^J(w)\sim -\eta^{IJ}\log(z-w),~~
\psi_L^I(z)\psi_L^J(w)\sim -{\eta^{IJ}\over z-w} \ ,
\end{equation}
it follows that
\begin{equation}
G_L^{\pm}(z)V_L^{(-1,-1)}(0)\sim {1\over z}\left(
V_L^{(0,-1)}(0),\,V_L^{(-1,0)}(0)\right) \ ,
\end{equation}
from which
\begin{eqnarray}
V_L^{(0,-1)}(z)={1\over2}(kJ_-^L\psi_L(z))\,e^{ik\cdot x_L(z)},\quad
V_L^{(-1,0)}(z)={1\over2}(kJ_+^L\psi_L(z))\,e^{ik\cdot x_L(z)} \ .
\end{eqnarray}
Here $J_{\pm}^L=\eta\pm i\cJ^L$. 
Also
\begin{equation}
G_R^{\pm}(\bar{z})V_R^{(-1,-1)}(0)\sim {1\over \bar{z}}\left(
V_R^{(0,-1)}(0),\,V_R^{(-1,0)}(0)\right) \ ,
\end{equation}
from which
\begin{eqnarray}
V_R^{(0,-1)}(\bar{z})={1\over2}(kJ_-^R\psi_R(\bar{z}))\,
e^{ik\cdot x_R(\bar{z})},\quad
V_R^{(-1,0)}(\bar{z})={1\over2}(kJ_+^R\psi_R(\bar{z}))\,
e^{ik\cdot x_R(\bar{z})}.
\end{eqnarray}
Here we define $J_{\pm}^R=\eta\pm i\cJ^R$.
Finally, the vertex operator of $(0,0)$-picture is
\begin{equation}
G_L^-(z)V_L^{(0,-1)}(0)-G_L^+(z)V_L^{(-1,0)}(0)\sim
{1\over z}\,V_L^{(0,0)}(0),
\end{equation}
from which
\begin{equation}
V_L^{(0,0)}(z)=\left( -k\cJ^L\partial x_L+
{1\over 2}\left(kJ^L_+\psi_L\right)\left(kJ^L_-\psi_L\right)\right)
e^{ik\cdot x_L} \ .
\end{equation}
The right sector vertex operator of the same picture takes the same form
with $L\rightarrow R$.

As discussed in \cite{ov}, the scalar $\phi$ of the $\beta$-string is
a deformation of the K\"ahler structure (\ref{k0;beta})
as $K_0 \rightarrow K_0 + \phi$. 
What is the interpretation of the scalar  $\phi$ of the
$\alpha$-string?
A generating function for the scattering amplitudes is
\begin{equation}
\Big\langle\exp\left(\int
  d^2zd^4\theta\,\phi(X^1,\bar{X^1},\Xt^2, \Xtb )\right)\Big\rangle \ ,
\end{equation}
where
\begin{equation}
\phi(X^1,\bar{X^1},\Xt^2, \Xtb)=\int d^4k \,\widetilde{\phi}(k)
 V_{\alpha}(k) ,
\end{equation}
and $V_{\alpha}(k)$ is the scalar vertex operator (\ref{vertex;alpha}). 
Thus, $\phi$ is a  deformation of $\hat{K}_0$ in (\ref{lt})
$\hat{K}_0\rightarrow \hat{K}_0 +\phi$.

In the next section, we will compute the scattering amplitudes of
$\alpha$-string and derive the effective action for $\phi$.

\section{Scattering Amplitudes}

In this section, we compute the three and four point amplitudes
of $N=2$ strings on a sphere.
Let us begin with the three point amplitude.
Recall that the total superconformal ghost number of the amplitude
on the sphere
is two.
The amplitude takes the form
\begin{equation}
\cA_3=\cA_3^L\,\cA_3^R,
\end{equation}
where $\cA_3^L$ ($\cA_3^R$) is the left (right) sector contribution
\begin{eqnarray}
\cA_3^L&\!=\!&
\langle V^{(-1,-1)}_L(z_1)\,V^{(-1,-1)}_L(z_2)\,V^{(0,0)}_L(z_3)\rangle \nn
&&\cdot\langle c_L(z_1)\,c_L(z_2)\,c_L(z_3)\rangle\,
\langle e^{-\vphi_L^+(z_1)}\,e^{-\vphi_L^+(z_2)}\rangle\,
\langle e^{-\vphi_L^-(z_1)}\,e^{-\vphi_L^-(z_2)}\rangle \ . 
\end{eqnarray}
Here $c_L$ is the  spin $-1$ 
conformal ghost, and $\vphi^{\pm}$
come from the bosonization of the two sets of superconformal ghosts
$\beta^{\pm},\gamma^{\pm}$.
To evaluate the ghost part in $\cA_3$, we use the formulas
\begin{eqnarray}
\langle c_L(z_1)c_L(z_2)c_L(z_3)\rangle&\!=\!&(z_1-z_2)(z_2-z_3)(z_1-z_3),
\nn
\left\langle e^{-\vphi_L^{\pm}(z_1)}e^{-\vphi_L^{\pm}(z_2)}\right\rangle
&\!=\!&{1\over z_1-z_2} \ .
\end{eqnarray}
By setting, say, $z_1=\infty,~z_2=0,~z_3=1$, we obtain
\begin{equation}
{\cal A}_3=c^L_{ij}\,c^R_{ij} \ ,
\end{equation}
where
\begin{equation}
c_{ij}^{L(R)}=-ik_i^I\cJ_{IJ}^{L(R)}k_j^J \ ,
\end{equation}
and $i\ne j$.
For $\beta$-string, $\cJ^L=\cJ^R=\cJ$ so that
\begin{equation}
{\cal A}_3^{\beta}=(c_{ij})^2,
\label{3pt;beta}
\end{equation}
where 
\begin{equation}
c_{ij}=-ik_i^I\cJ_{IJ}k_j^J \ ,
\end{equation}
as in (\ref{cij}) and (\ref{cijn}).
This reproduces the result of \cite{ov}.

For $\alpha$-string,
$\cJ^L=\cJ,\cJ^R=\widetilde{\cJ}$ and
\begin{equation}
{\cal A}_3^{\alpha}=c_{ij}\,\ct_{ij} \ ,
\label{3pt;alpha}
\end{equation}
where 
\begin{equation}
\widetilde{c}_{ij}=-ik_i^I\widetilde{\cJ}_{IJ}k_j^J \ ,
\end{equation}
as in (\ref{ctij}).

The 
two amplitudes ${\cal A}_3^{\beta}$ and ${\cal A}_3^{\alpha}$
are related by T-duality
along one of two spatial directions \footnote{Under T-duality along a 
time-like direction the amplitudes are equal up to a sign.}.
Consider, for instance,  a T-duality along the $x^4$ direction, which is
implemented by $k^4_R\rightarrow -k^4_R$. Here $k_R$ is the momentum of
a string in the right sector.
It is easy to see that 
\begin{equation}
c_{ij}(k^1,k^2,k^3,-k^4)=\ct_{ij}(k^1,k^2,k^3,+k^4) \ .
\end{equation}

The amplitude (\ref{3pt;beta}) is reproduced by an effective action
for the scalar \cite{ov} satisfying
the Plebanski equation \cite{P}.
We turn to the amplitude (\ref{3pt;alpha}) and ask what
is the effective action for the scalar that reproduces it.
Here we find a somewhat surprising result. 

Consider three on-shell momenta $k_i$ obeying
$k_1+k_2+k_3=0$.
It is useful to work in the complex basis of section 2, where
the on-shell momenta take the form $k=(k^l,\bar{k}^{\bar{l}})$ with
$k^1_i=k_ie^{i\theta_i},k^2_i=k_ie^{i\phi_i}$ and
$k_i,\theta_i,\phi_i$ are real.
Momentum conservation implies the relation
\begin{equation}
\cos(\theta_i-\theta_j)=\cos(\phi_i-\phi_j) \ .
\end{equation}
We can recast the amplitude as
\begin{equation}
c_{ij}\ct_{ij}=4(k_ik_j)^2\left[ 
\cos^2(\theta_i-\theta_j)-\cos^2(\phi_i-\phi_j)\right] \ ,
\end{equation}
and hence $\cA_3^{\alpha}=0$. This means that the effective action
of the scalar 
of the $\alpha$-string has no three-point interaction.

Let us turn now to the computation of the four-point amplitude.
Consider
\begin{equation}
\cA_4=\int d^2z\cA_4^L\,\cA_4^R \ ,
\end{equation}
where $\cA_4^L$ comes from the left sector
\begin{eqnarray}
\cA_4^L&\!=\!&
\langle V^{(-1,-1)}_L(z_1)\,V^{(-1,-1)}_L(z_2)\,V^{(0,0)}_L(z_3)
        \,V^{(0,0)}_L(z)\rangle\, \nn
&&\langle c_L(z_1)\,c_L(z_2)\,c_L(z_3)\rangle\,
\langle e^{-\vphi_L^+(z_1)}\,e^{-\vphi_L^+(z_2)}\rangle\,
\langle e^{-\vphi_L^-(z_1)}\,e^{-\vphi_L^-(z_2)}\rangle  \ , 
\end{eqnarray}
and a similar expression for the right sector.
To evaluate the amplitude, we need the following formulas that involve
the complex structures
\begin{equation}
J_{\pm}\eta^{-1}J_{\pm}=2J_{\pm},~~J_{\pm}\eta^{-1}J_{\mp}=0 \ ,
\end{equation}
where $J_{\pm}=\eta\pm i\cJ$,
and the integral formula
\begin{eqnarray}
&&\int d^2z\,z^{\alpha+n_1}\bar{z}^{\alpha+n_2}
(1-z)^{\beta+m_1}(1-\bar{z})^{\beta+m_2} \nn
&&={\sin(\pi\alpha)\sin(\pi\beta)\over\sin(\pi(\alpha+\beta))}\,
{\Gamma(\alpha+n_1+1)\,\Gamma(\alpha+n_2+1)
\,\Gamma(\beta+m_1+1)\,\Gamma(\beta+m_2+1) \over
\Gamma(\alpha+\beta+n_1+m_1+2)\,\Gamma(\alpha+\beta+n_2+m_2+2)} \ .
\end{eqnarray}
Setting, say, $z_1=1,~z_2=\infty,~z_3=0$,
it is straightforward to show that 
\begin{equation}
\cA_4=
\pi\,F^LF^R\,
\,{\Gamma(1-s)\,\Gamma(1-t)\, \Gamma(1-u)\over\Gamma(s)\, \Gamma(t)\, 
\Gamma(u)} \ ,
\end{equation}
with 
\begin{equation}
F^{L(R)}=1-{c_{12}^{L(R)}c_{34}^{L(R)}\over su}
-{c_{23}^{L(R)}c_{41}^{L(R)}\over tu} \ .
\end{equation}
Here $s=-k_1\cdot k_2,~t=-k_2\cdot k_3,~u=-k_1\cdot k_3$.
%
Using the (\ref{id;c}) 
we find that 
\begin{equation}
\cA_4^{\alpha}=\cA_4^{\beta}=0.
\end{equation}

Consider now the higher-point amplitudes.
As discussed in \cite{ov}, 
these amplitudes, if nonzero, will have infinite
number of poles corresponding to unphysical massive string states, and
therefore should vanish.
Alternatively, 
it has been  argued 
in \cite{BV,Ber} that for $N=2$ strings in
a flat background, all the scattering amplitudes vanish except
the sphere three-point amplitude.
This stronger claim implies that the $\alpha$-string scalar is
free to all orders in string perturbation theory.
Note that, as argued in \cite{Lechtenfeld:1997za}, the inclusion of contributions from sectors of 
nonzero worldsheet $U(1)$ instanton numbers
does not modify the result.

\section{Effective Action and Geometry}

In this section we discuss the effective action and geometry of the 
$N=2$ strings. 
As shown in \cite{ov}, the effective action of the $\beta$-string
scalar is given by 
\begin{equation}
\label{P}
S_{\beta}=\int\left(\partial\phi\bar{\partial}\phi
+{1\over 3}\,\phi\,\partial\bar{\partial}\phi\wedge
\partial\bar{\partial}\phi\right) \ .
\end{equation}
This effective action reproduces
the correct three-point and four-point amplitudes of the
string computation.
It has not been verified yet that all the
higher-order amplitudes arising from (\ref{P}) vanish \footnote{For partial results in the open 
and closed strings cases see \cite{Parkes:1992rz}.}.

Recall that $\phi$ is a deformation of the flat space  K\"ahler potential.
The field equation for $\phi$ is the Ricci flatness condition.
Thus, the background is  K\"ahler and  Ricci-flat, which in four
dimensions is equivalent to the curvature being self-dual.

Consider next 
the $\alpha$-string scalar.
We have seen that all the $n(\ge3)$-point amplitudes
vanish. This implies that $\alpha$-string scalar
is free and the effective action is 
\begin{equation}
S_{\alpha}=\int\partial\phi\bar{\partial}\phi \ .
\label{ea;alpha}
\end{equation}

We have seen that $\phi$ is a deformation of the potential
$\widehat{K}_0$ in (\ref{khat}), so that the $\sigma$-model action
is
\begin{equation}
\label{action}
S=\int d^2\sigma d^4\theta \,\Kh(X,\bar{X},\Xt,\bar{\Xt}) \ ,
\end{equation}
where $\Kh=\widehat{K}_0 +\phi$.
The geometry of the theories described by (\ref{action})
has been studied in \cite{GHR}.
Define
\begin{eqnarray}
\label{metriceq}
&&g_{X\bar{X}}=\partial_X\partial_{\bar{X}}\Kh,~~
g_{\Xt\bar{\Xt}}=-\partial_{\Xt}\partial_{\bar{\Xt}}\Kh,\nn
&&B_{X\bar{\Xt}}=\partial_X\partial_{\bar{\Xt}}\Kh,~~
B_{\Xt\bar{X}}=\partial_{\Xt}\partial_{\bar{X}}\Kh \ ,
\end{eqnarray}
then by examining the scalar component of the action one finds that
$g_{IJ}$ is the metric of the target space and
$B_{IJ}$ is a two-form field.

One defines a connection with torsion by
\begin{equation}
\label{tor}
\Gamma^{\pm K}{}_{IJ}=\Gamma^{K}{}_{IJ}\mp
H^{K}_{IJ} \ ,
\end{equation}
where $H = dB$.
Conformal invariance requires 
the self-duality condition of the Riemann tensor
with torsion (\ref{tor}) $R^{\pm}_{IJKL}$ \cite{Hull}
\begin{equation}
R^{\pm}_{IJKL}={1\over 2}\epsilon_{KLMN}
R_{IJ}^{\pm}{}^{MN} \ ,
\end{equation}
and
\begin{equation}
R^+_{IJKL}=R^-_{KLIJ} \ .
\end{equation}

The derivation of the result that  $\Kh$ is a free scalar
using the worldsheet computations that we performed 
is in agreement
with \cite{Hull} that predicted this based on the conformal invariance
of the $\sigma$-model.
We note that while
in the $\sigma$-model the result is expected to hold
to all orders in
$\alpha^{\prime}$, here we expect that it 
holds to all orders 
in the string loop expansion.

\section{An $\alpha$-string Background}

As an example we show that the four-dimensional transverse part of
an NS5-brane background \cite{chs} solves the field equations of the
$\alpha$-string. This is expected since the  NS5-branes have been
argued to be  T-dual
to ALE spaces 
\cite{Ooguri:1995wj}, which are solutions of the $\beta$-string
field
equations. Thus, by T-duality we expect the NS5-branes to solve 
the $\alpha$-string field equations.

We consider first the signature $(4,0)$.
The transverse part to the NS5-branes background reads \cite{chs}
\begin{eqnarray}
&&ds^2=e^{2\Phi}\delta_{\mu\nu}dx^{\mu}dx^{\nu} \ ,\nn
&&H=dB=2Q\epsilon_3 \ ,\nn
&&e^{2\Phi}=e^{2\Phi_0}+{Q\over x^2} \ .
\label{ns5}
\end{eqnarray}
$\mu,\nu = 0,...,3$ and $x^2=x_{\mu}x^{\mu}$.
$\epsilon_3$ is the
volume form of a unit $\bS^3$.

We look for a potential $\phi$ that reproduces this
background with $\mu_p=0$.
Thus, $\phi$ obeys the free field equation in four
dimensions.
Using the metric equations (\ref{metriceq}) we obtain
\begin{equation}
\partial_1\partial_{\bar{1}}\phi=-\partial_2\partial_{\bar{2}}\phi
=e^{2\Phi_0}-1+{Q\over x^2} \ .
\end{equation}
 To solve this, we assume that $\phi$ depends on
$r_1=|z_1|$ and $r_2=|z_2|$.
$\phi$ takes the form
\begin{eqnarray}
\phi&\!=\!&{e^{2\Phi_0}-1\over 4}(r_1^2-r_2^2)
-{Q\over 4}\,{\rm Li}_2\left(-{r_1^2\over r_2^2}\right)+Q\log r_1\log r_2
-{Q\over 2}\left(\log r_2\right)^2 \nn &&
+(k_1+k_2\log r_1)(k_3+k_4\log r_2)  \ .
\end{eqnarray}
Here $k_1,k_2,k_3,k_4$ are constant and ${\rm Li}_2(x)$ is the
polylogarithm function satisfying
\begin{equation}
{d\over dx}{\rm Li}_2(x)=-{\log(1-x)\over x} \ .
\end{equation}
Note that the potential depends on four free parameters.
These correspond to symmetry  transformations that leave
the background unchanged.
This is similar to K\"ahler transformations of
$\phi$ in the $\beta$-string.

The solution with signature $(2,2)$ is obtained by $r_1\rightarrow ir_1$.

\section*{Acknowledgements}

Y.O. would like to thank Z. Yin for valuable discussions. 
This research is supported by the US-Israel Binational Science
Foundation.

\newpage



\begin{thebibliography}{99}

\bibitem{ademollo;76}
M.~Ademollo {\it et al.},
``Dual String With U(1) Color Symmetry,''
Nucl.\ Phys.\ B {\bf 111}, 77 (1976).


\bibitem{ov}
H.~Ooguri and C.~Vafa,
``Selfduality And N=2 String Magic,''
Mod.\ Phys.\ Lett.\ A {\bf 5}, 1389 (1990);
H.~Ooguri and C.~Vafa,
``Geometry of N=2 strings,''
Nucl.\ Phys.\ B {\bf 361}, 469 (1991).





\bibitem{oz;n2}
Y.~K.~Cheung, Y.~Oz and Z.~Yin,
``Families of N = 2 strings,''
arXiv:hep-th/0211147.


\bibitem{Gates:1988tn}
S.~J.~Gates, L.~Lu and R.~N.~Oerter,
``Simplified SU(2) Spinning String Superspace Supergravity,''
Phys.\ Lett.\ B {\bf 218} (1989) 33.



\bibitem{GHR}
S.~J.~Gates, C.~M.~Hull and M.~Rocek,
``Twisted Multiplets And New Supersymmetric Nonlinear Sigma Models,''
Nucl.\ Phys.\ B {\bf 248}, 157 (1984).


\bibitem{Hull}
C.~M.~Hull,
``The geometry of N = 2 strings with torsion,''
Phys.\ Lett.\ B {\bf 387}, 497 (1996)
[arXiv:hep-th/9606190].



\bibitem{Buscher:sk}
T.~H.~Buscher,
``A Symmetry Of The String Background Field Equations,''
Phys.\ Lett.\ B {\bf 194} (1987) 59.


\bibitem{Buscher:uw}
T.~Buscher, U.~Lindstrom and M.~Rocek,
``New Supersymmetric Sigma Models With Wess-Zumino Terms,''
Phys.\ Lett.\ B {\bf 202}, 94 (1988).


\bibitem{Ivanov:ec}
I.~T.~Ivanov, B.~b.~Kim and M.~Rocek,
``Complex Structures, Duality And WZW Models In Extended Superspace,''
Phys.\ Lett.\ B {\bf 343}, 133 (1995)
[arXiv:hep-th/9406063].


\bibitem{FT}
E.~S.~Fradkin and A.~A.~Tseytlin,
``Quantum String Theory Effective Action,''
Nucl.\ Phys.\ B {\bf 261} (1985) 1.



\bibitem{fms}
D.~Friedan, E.~J.~Martinec and S.~H.~Shenker,
``Conformal Invariance, Supersymmetry And String Theory,''
Nucl.\ Phys.\ B {\bf 271}, 93 (1986).

\bibitem{Li:1992rr}
M.~Li,
``Gauge symmetries and amplitudes in N=2 strings,''
Nucl.\ Phys.\ B {\bf 395} (1993) 129
[arXiv:hep-th/9204027];
J.~Bischoff, S.~V.~Ketov and O.~Lechtenfeld,
``The GSO projection, BRST cohomology and picture changing in N=2 string theory,''
Nucl.\ Phys.\ B {\bf 438} (1995) 373
[arXiv:hep-th/9406101].

\bibitem{P}
J.~F.~Plebanski,
``Some Solutions Of Complex Einstein Equations,''
J.\ Math.\ Phys.\  {\bf 16}, 2395 (1975).





\bibitem{BV}
N.~Berkovits and C.~Vafa,
``N=4 topological strings,''
Nucl.\ Phys.\ B {\bf 433}, 123 (1995)
[arXiv:hep-th/9407190].


\bibitem{Ber}
N.~Berkovits,
``Vanishing theorems for the selfdual N=2 string,''
Phys.\ Lett.\ B {\bf 350}, 28 (1995)
[arXiv:hep-th/9412179].

\bibitem{Lechtenfeld:1997za}
O.~Lechtenfeld and W.~Siegel,
``N = 2 worldsheet instantons yield cubic self-dual Yang-Mills,''
Phys.\ Lett.\ B {\bf 405} (1997) 49
[arXiv:hep-th/9704076].

\bibitem{Parkes:1992rz}
A.~Parkes,
``A Cubic action for selfdual Yang-Mills,''
Phys.\ Lett.\ B {\bf 286} (1992) 265
[arXiv:hep-th/9203074];
G.~Chalmers, O.~Lechtenfeld and B.~Niemeyer,
``N = 2 quantum string scattering,''
Nucl.\ Phys.\ B {\bf 591} (2000) 39
[arXiv:hep-th/0007020].


\bibitem{chs}
C.~G.~Callan, J.~A.~Harvey and A.~Strominger,
``Worldbrane actions for string solitons,''
Nucl.\ Phys.\ B {\bf 367}, 60 (1991).



\bibitem{Ooguri:1995wj}
H.~Ooguri and C.~Vafa,
``Two-Dimensional Black Hole and Singularities of CY Manifolds,''
Nucl.\ Phys.\ B {\bf 463} (1996) 55
[arXiv:hep-th/9511164].






\end{thebibliography}
\end{document}